\documentclass[]{aastex701}
\begin{document}

\title{Supergranulation as a Tracer of Solar-Cycle Variability}

\author[orcid=0000-0003-4144-2270]{Irina N. Kitiashvili}
%\altaffiliation{}
\affiliation{Computational Physics Branch, NASA Advanced Supercomputing Division, NASA Ames Research Center}
\email[show]{irina.n.kitiashvili@nasa.gov}  

\author[orcid=0009-0001-9280-6487]{Andrew A. Ngo} 
\affiliation{Department of Mathematical Sciences, University of Delaware}
\email{ango@udel.edu}

\author[orcid=0000-0002-0972-8642]{Spiridon Kasapis}
\affiliation{Department of Astrophysical Sciences, Princeton University}
\email{skasapis@princeton.edu}

\begin{abstract}
Supergranulation is a dominant scale of near-surface solar convection and an important diagnostic for studying the interaction between convective flows, rotation, and magnetic activity. We develop an automated framework to identify and characterize supergranular structures using subsurface horizontal-velocity maps inferred from time-distance helioseismology. The method applies Laplacian-of-Gaussian detection to horizontal velocity divergence maps and combines it with magnetic-field masking, divergence-based filtering, and multi-peak rejection to isolate supergranular cells under both quiet-Sun and active conditions.

Using this approach, we analyze 16 years of SDO/HMI observations spanning Solar Cycles 24 and 25 and investigate temporal and latitudinal variations of supergranular properties. The results reveal a clear solar-cycle dependence: the mean supergranular diameter decreases near solar maxima and increases around solar minima, with systematically larger supergranules in the weaker Solar Cycle 24 than in the stronger Solar Cycle 25. The width of the diameter distribution increases during high magnetic activity, whereas the number of detected supergranules shows only weak cycle dependence. The mean velocity divergence within detected supergranules is strongly anticorrelated with sunspot number, indicating magnetic suppression of divergent convective flows. Time-latitude distributions of supergranular diameter and divergence closely follow the migration of solar activity belts, while the curl of the horizontal velocity shows weak cycle dependence but pronounced hemispheric asymmetry. These results establish supergranulation as a sensitive tracer of solar-cycle-related changes in near-surface convection.
\end{abstract}

\section{Introduction}\label{Introduction} 

Supergranulation is one of the most prominent manifestations of solar convection, covering almost the entire solar surface and characterized by flow patterns with typical scales of 20–30~Mm and lifetimes of about a day \citep[e.g.,][]{Leighton1962,Simon1964,Rincon2018}. These flows play a crucial role in organizing the magnetic field on the solar surface, influencing flux transport and network formation, and shaping the solar atmosphere from the chromosphere to the heliosphere. Therefore, understanding the origin, structure, and depth extent of supergranulation, as well as its role in energy transport and coupling with solar activity across multiple scales, is important for both fundamental physics and improving solar activity forecasting.
In addition, it is known that the signal from supergranulation is a significant contributor to the star-induced radial velocity variations known as `stellar jitter' \citep[e.g.,][]{Meunier2020}. Therefore, changes in the properties of supergranulation associated with global variability of a star might add extra modulation to the radial velocity signal \citep{OSullivan2025}.

Typically, supergranulation patterns are identified by analyzing the distribution of small-scale magnetic fields, which accumulate at cell boundaries; Ca II K observations, where these magnetic concentrations appear as localized brightenings; time-averaged Doppler velocity maps, which suppress granulation signals; or horizontal flow divergence maps (Table~\ref{tab:SG-obs}).
Because supergranulation rotates about 3\% faster than the photosphere \citep[e.g.,][]{Duvall1980,Beck2000}, it is believed to be rooted 10--20~Mm below the solar surface. This connection to subsurface layers makes supergranulation an important diagnostic of solar convection and magnetic activity. The influence of solar activity has been explored by many authors. Previous studies can be broadly divided into two groups (Table~\ref{tab:SG-obs}): those that consider the properties of convective patterns within a specific area of the Sun over a short period of time, and those spanning at least part of a solar cycle. In these studies, supergranulation has been identified using a variety of observables (e.g., Doppler velocity maps, magnetograms, and Ca II K images) and analysis methods (e.g., autocorrelation and gradient-based techniques). This lack of consensus suggests an incomplete understanding of how convection responds to varying magnetic flux levels on spatial scales significantly larger than granulation, which may contribute to the diversity of reported results.

In this paper, we investigate the relationship between global solar activity and the properties of supergranulation by analyzing convective flows over Solar Cycles 24 and 25 observed with the Helioseismic and Magnetic Imager on board the Solar Dynamics Observatory \citep[SDO/HMI;][]{Scherrer2012}. Supergranules are identified using a Laplacian-of-Gaussian (LoG) blob detection method \citep{Lindeberg1998} applied to horizontal velocity divergence maps. We begin by describing the observations and initial data reduction procedures (Section~\ref{sec:obs}), then provide an overview of the data analysis pipeline (Section~\ref{sec:pipeline}) and summarize the adopted parameters, along with examples of detection performance (Section~\ref{sec:performance}). We then present the results based on approximately 16 years of SDO/HMI observations (Section~\ref{sec:results}), followed by a discussion of the main findings (Section~\ref{sec:discussion}) and conclusions (Section~\ref{sec:conclusion}).

\section{Observations and Initial Data Reduction}\label{sec:obs}

To identify supergranulation patterns and track their evolution over Solar Cycles 24 and 25, we utilize time-distance helioseismology data products, available through the Joint Science Operations Center (JSOC) Science Data Processing (SDP) portal\footnote[1]{\url{http://jsoc.stanford.edu/}}. The data products are reconstructed from $5\times5$ overlapping patches, tracked with the Carrington rotation rate, and reprojected using Postel's projection. The inference of the subsurface flow field is obtained from the Doppler velocity using the Gabor wavelet method and combined into $120^{\mathrm{o}}\times120^{\mathrm{o}}$ maps with a spatial sampling of $0.117^{\mathrm{o}}$ pixel$^{-1}$ and cadence of 8 hours for six depth ranges covering layers from the photosphere to 35~Mm below~\citep{Zhao2012}. 

\begin{figure}[b]
\centerline{\includegraphics[width=1\textwidth,clip=]{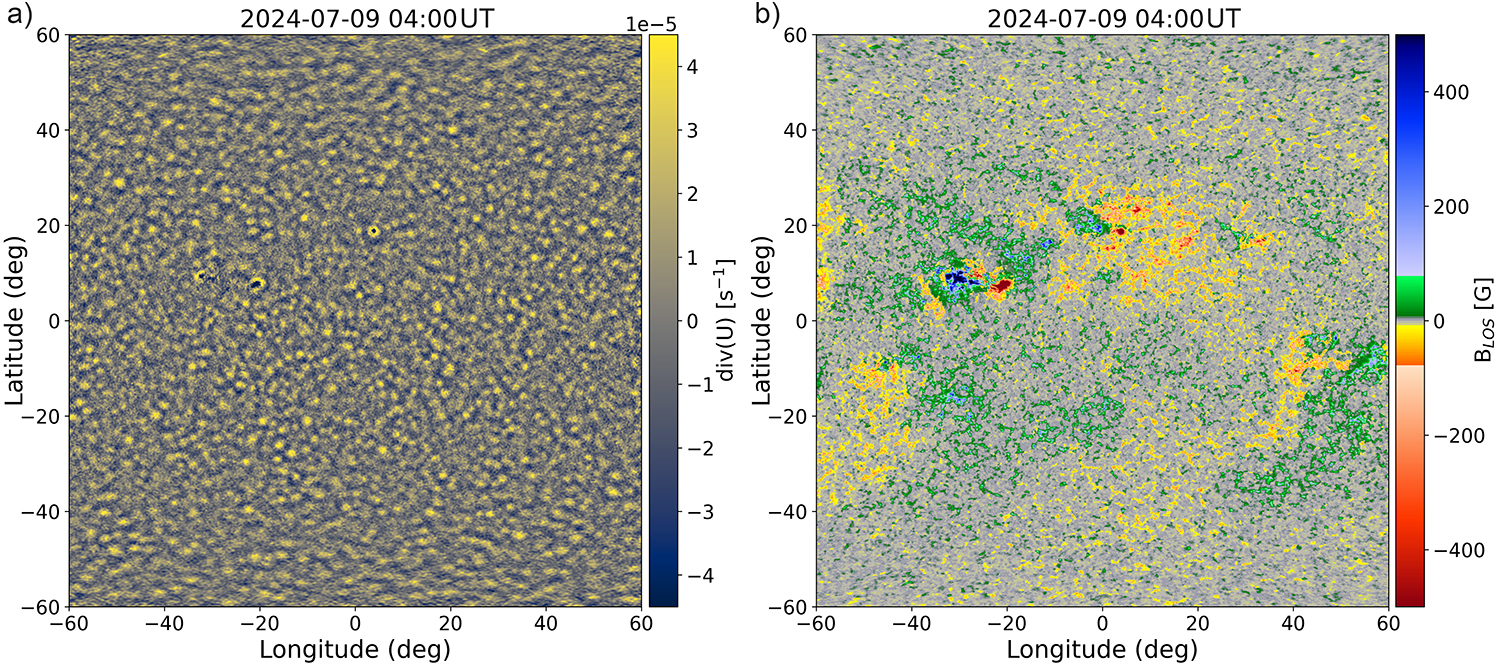}}
\caption{Velocity divergence map (panel a) and corresponding the line-of-sight magnetogram (panel b).}
\label{fig:divUB}
\end{figure}

In this work, we used the horizontal flow maps obtained from 1st May 2010 to 14 March 2026, taken at 04:00~UT, assuming a 24-hour characteristic lifetime for supergranulation \citep{Rincon2018}. We identified five time points at which only one component of the horizontal velocity is available. These data gaps have been replaced with data from the same day at 12:00 UT to ensure consistency in the analysis. Following \cite{Zhao2012}, we use flow maps corresponding to the 1--3~Mm layer below the photosphere to compute velocity divergence (Figure~\ref{fig:divUB}a). The annual variations in pixel size caused by changes in Earth's distance from the Sun were accounted for. As expected, the supergranulation patterns (yellow localized circular or stretched instances at high latitudes in Figure~\ref{fig:divUB}a) are the most populated features. 

To reduce foreshortening and projection-related distortions, we restrict the analysis to selected latitude ranges. Because the inferred supergranular properties may depend on the latitude range considered, we present the results for three ranges: 1) $\pm 30^{\mathrm{o}}$, 2) $\pm 45^{\mathrm{o}}$, and 3) $\pm 60^{\mathrm{o}}$, corresponding to the full latitude extent of the available data.

\section{Data Processing Pipeline}\label{sec:pipeline}

To perform a robust identification of supergranules over the whole period of SDO/HMI observations, we developed a unified data processing pipeline that includes 1) noise reduction, 2) diagnostic analysis, 3) supergranule detection, and 4) additional filtering to remove irrelevant structures.

\subsection{Gaussian Denoising}\label{sec:denoising}
\begin{figure}[t]
\centerline{\includegraphics[width=0.95\textwidth,clip=]{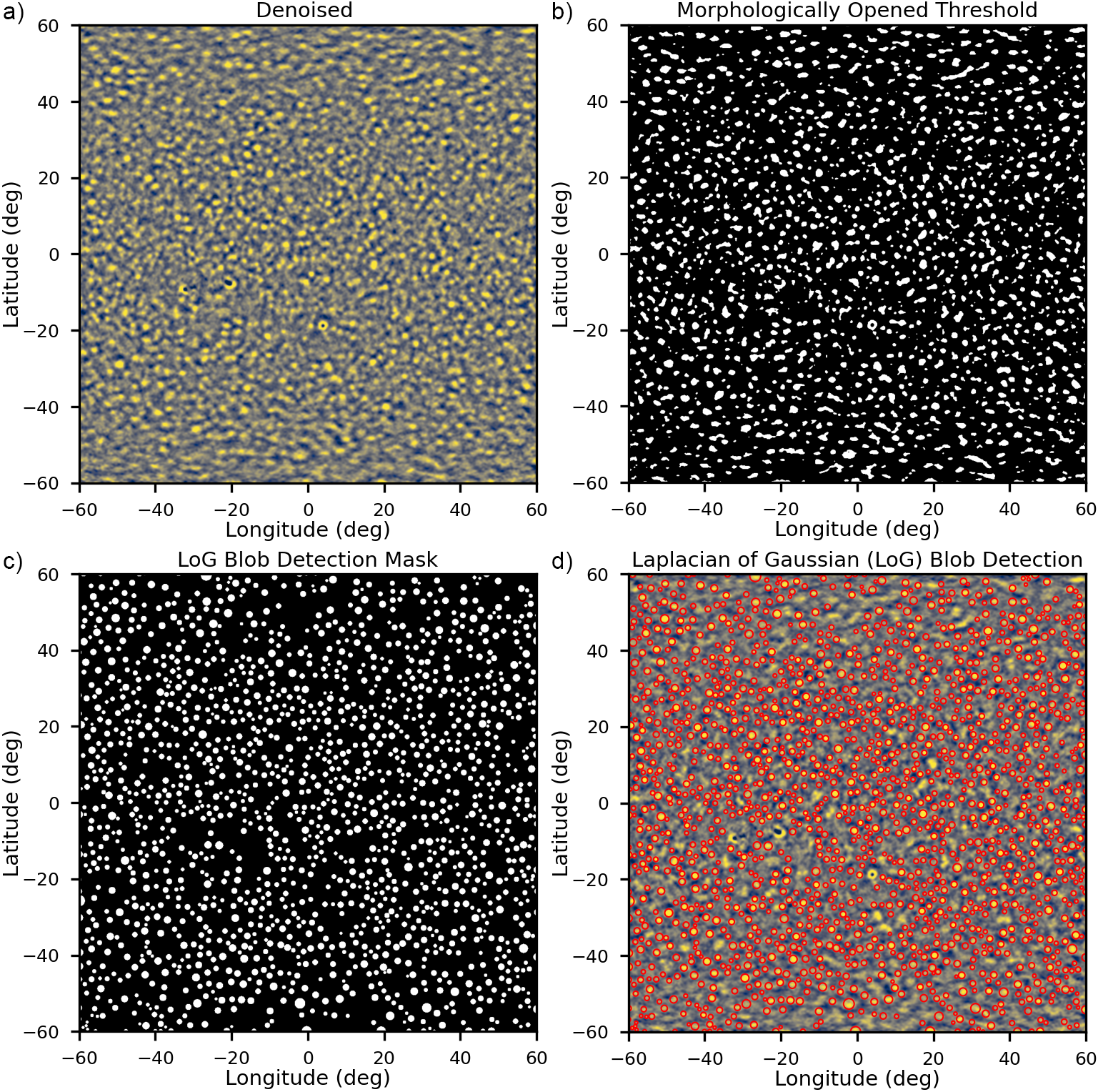}}
\caption{Example of the data processing pipeline at different stages applied to a horizontal velocity divergence map for 9 July, 2024, during high solar activity. Panel a) shows the divergence map after trimming and application of Gaussian denoising. Panel b) illustrates the result of thresholding and morphological opening performed for diagnostic purposes. Panels c) and d) show identified supergranules using the LoG algorithm after masking active regions and filtering.}
\label{fig:method}
\end{figure}

Several sources of noise affect the velocity divergence maps, including turbulent convection, granulation, time-distance inversion analysis, and the HMI processing pipeline. To suppress small-scale fluctuations, we smoothed each divergence map with a two-dimensional isotropic Gaussian kernel
\begin{equation}\label{eq:gauss-sm}
    G(\Delta x, \Delta y) = \frac{1}{2\pi\sigma_{\rm ph}^2}\exp\left(-\frac{\Delta x^2 + \Delta y^2}{2\sigma_{\rm ph}^2}\right),
\end{equation}
where $\sigma_{\rm ph}$ is the standard deviation of the Gaussian smoothing kernel, expressed in physical units of Mm, and $\Delta x$ and $\Delta y$ represent the distance from a pixel center. The kernel was applied as a sliding window across all pixels of the maps. In discrete form, the Gaussian width in pixels is $\sigma=\sigma_{\rm ph}/dx$, where $dx$ is the pixel size. In this work, we applied $\sigma_{\rm ph}=3$~Mm, which corresponds to the effective spatial scale $\lambda_{\rm eff}=\sigma_{\rm ph} \sqrt{2} \sim 4.24$~Mm (Figure~\ref{fig:method}a).

\begin{figure}[t]
\centerline{\includegraphics[width=1\textwidth,clip=]{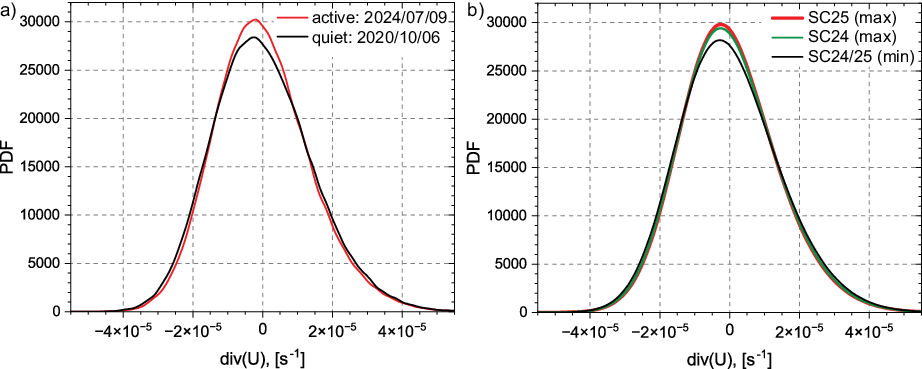}}
\caption{Probability density functions of the horizontal velocity divergence obtained for a) the most extreme magnetically quiet (6 October, 2020; black curve) and active (9 July, 2024; red curve) days, and different levels of the solar activity, averaged over 2 years: maximum of solar cycle 25 (1 January, 2023 -- 1 January, 2025; red curve), maximum of solar cycle 24 (1 January, 2013 -- 1 January, 2015; green curve), and minimum of the solar activity between these cycles (1 July, 2018 -- 1 July, 2020).}
\label{fig:hist_minmax}
\end{figure}

\subsection{Adaptive Thresholding and Morphological Opening}\label{sec:morph-op}

It is well known that during a solar cycle, the distribution and strength of magnetic flux vary significantly, affecting the properties of convective flows across a wide range of spatial scales. Taking into account the nonlinear coupling between magnetic fields and convective flows, we apply an adaptive threshold based on the distribution of the velocity divergence. To evaluate the distribution of the velocity divergence, we selected the moment when the Sun was most magnetically active over the entire observation period (9 July, 2024) and the moment when it was extremely quiet for several consecutive days (October, 2020) to minimize the impact of potential subsurface activity. In general, the distribution of the flow divergence, ${\rm div(U)}$, is close to Gaussian and exhibits variations with magnetic activity. During high solar activity, the distribution becomes more peaked and exhibits reduced variance, reflecting the suppression of horizontal convective flows in magnetically active regions (Figure~\ref{fig:hist_minmax}a). 
Two-year averages centered on solar maximum and minimum for Solar Cycles 24 and 25 demonstrate a clear dependence of the divergence distribution on magnetic activity level. In particular, the distribution becomes narrower during high-activity periods, consistent with suppression of convective flows in magnetized regions, with a weaker modulation observed during the comparatively low-amplitude Cycle 24 (Figure~\ref{fig:hist_minmax}b).

To account for temporal variations in the divergence distribution associated with changes in solar magnetic activity, we define an adaptive initial threshold as the 85th percentile of the divergence field for each map. The percentile level was initially estimated from the divergence distribution (Figure~\ref{fig:hist_minmax}b), where the largest positive divergence values are primarily associated with supergranular outflows. The threshold was subsequently calibrated using representative observations obtained during periods of minimum and maximum solar activity. The final percentile value was selected to maximize the detection rate of supergranules while minimizing noise-induced detections and preventing the fragmentation of individual supergranular cells into multiple detections.

This percentile-based threshold provides a robust, data-driven criterion for isolating regions of strong positive divergence. The percentile-based threshold does not exclude enhanced divergence associated with active regions (e.g., Figure~\ref{fig:divUB}) and is used only for diagnostic purposes. To further refine the spatial structure of these high-divergence regions, we apply a morphological opening operation \citep{Serra1982} to the thresholded maps, which removes small-scale noise and isolates coherent patches of enhanced divergence. Supergranule identification is instead performed using multi-scale Laplacian-of-Gaussian (LoG) detection applied to the normalized divergence field, followed by additional filtering criteria.

Since morphological opening is not widely used in solar observations, we briefly summarize its formal definition. In this context, the divergence maps ($D$) are treated as a grayscale image $f(x,y)$, to which morphological operations are applied. Morphological opening is defined as erosion followed by dilation \citep{Kimori2013}. The erosion operation removes small-scale structures and is defined as
\begin{equation}
        \varepsilon(f)_{(x,y)} = {\rm min}\{f(x+s, y+t)-b(s,t)| 
        (x+s,y+t) \in D_f; (s,t) \in D_b\},
\end{equation}
where $b(s,t)$ is a value at the offset pixels $(s,t)$, and $D_b$ denotes the set of discrete pixel offsets defining its shape. The operation is evaluated only for $(x+s,y+t)\in D_f$. The erosion operation assigns each pixel the minimum value within the neighborhood defined by the structuring element (kernel), reducing structure size and removing small-scale features. 

The complementary operation, dilation, expands structures by assigning to each pixel the maximum value within the neighborhood and is effective for filling small gaps, connecting nearby regions, and expanding structural boundaries \citep{Kimori2013}. It is defined as
\begin{equation}
        \delta(f)_{(x,y)} = {\rm max}\{f(x-s, y-t)+b(s,t)| 
        (x-s,y-t) \in D_f; (s, t) \in D_b\}.
\end{equation}

In this study, after applying thresholding, the divergence maps are represented in binary form ($f(x,y) \in \{0,1\}$); therefore, $b(s,t)=0$ within the kernel and is undefined outside it. In this case, the operation reduces to computing the minimum and maximum values over the neighborhood defined by the structuring element. This process removes small artifacts while preserving the shape and extent of larger, coherent regions of enhanced divergence. Although the morphologically opened image reveals the approximate spatial organization of supergranular patterns, the resulting regions are often clustered and do not uniquely correspond to individual supergranules (Figure~\ref{fig:method}b). This limitation is addressed in the subsequent detection step.

\subsection{Supergranule Identification with Laplacian-of-Gaussian Algorithm}

To identify individual supergranules, we employ a multi-scale blob detection approach based on the Laplacian-of-Gaussian (LoG) operator. The method follows the theoretical formulation of \cite{Marr1980} and the scale-space framework of \cite{Lindeberg1994}, enabling robust detection of spatially localized features while suppressing high-frequency noise. To account for the range of supergranular scales in the horizontal velocity divergence maps, we adopt a scale-space representation, in which the characteristic size of a detected structure is proportional to the Gaussian smoothing scale \citep{Lindeberg1998}.

In practice, this approach is implemented using the \texttt{blob-log} algorithm from the \texttt{scikit-image} package \citep{van2014scikit}. Unlike the morphologically opened binary representation (Section~\ref{sec:morph-op}), which often produces extended or clustered regions of enhanced divergence, the LoG method enables the identification of localized, scale-dependent structures corresponding to individual supergranular cells. However, the LoG detection algorithm may also capture high-divergence signals that are not associated with supergranulation patterns. To reduce such misidentification, additional filtering criteria are applied. Because this detection approach is relatively new in this context, we provide a brief description of the algorithm in Appendix~\ref{sec:AppendixB}.

\subsection{Post-processing and Filtering}

The initial detection with the LoG algorithm may identify spurious features, including noise-driven fluctuations and divergence enhancement associated with strong magnetic activity (e.g., Figure~\ref{fig:divUB}). To ensure the detected structures correspond to supergranular cells, additional post-processing and filtering steps are applied. These include thresholds applied to the physical divergence field to retain only regions with positive divergence that exceed a prescribed statistical significance level, excluding weak positive-divergence fluctuations unlikely to correspond to supergranular outflows. 

To remove detections associated with strong magnetic-field regions, the corresponding magnetogram is mapped onto the same spatial grid as the divergence field, enabling construction of an active-region mask. The mask is constructed by selecting pixels with magnetic field strength exceeding the threshold, $|B| \ge B_{\rm thr}=1000$~G. To account for the spatial extent of active regions, this mask is expanded using morphological dilation with a circular kernel of radius 20~px. Any LoG candidate whose center lies within the dilated magnetic mask is excluded from further analysis. 
After masking active regions, we applied a threshold to the divergence field to retain only statistically significant positive divergence regions. The threshold was defined as
\begin{equation}
D_{\rm thr}={\rm median}(D) + k \sigma_{\rm noise}, \label{eq:thr}
\end{equation}
where $k$ is a tunable parameter and $\sigma_{\rm noise}$ is a robust estimate of the background fluctuations computed from the median absolute deviation (MAD), $\sigma_{\rm noise} = 1.4826 \times {\rm median}\left(|D - {\rm median}(D)|\right)$. 
This approach reduces the influence of strong outliers and provides a stable noise estimate, allowing consistent identification of significant divergence features across the dataset.

To ensure each detected structure corresponds to a single supergranular cell, structures containing multiple local maxima have been removed. This is achieved by identifying secondary peaks within each LoG-defined region using a relative amplitude threshold and a minimum separation criterion of 3 pixels. Detections containing more than one significant peak within a cell, where the secondary peak exceeds a prescribed fraction of the primary peak amplitude, are removed from the analysis. The resulting set of detected supergranules is shown in Figure~\ref{fig:method}c, which provides a large sample of identified supergranules and effectively discriminates fragmented structures. A comparison between the detected supergranules and the underlying velocity divergence map reveals that some structures are not captured by the automated procedure (Figure~\ref{fig:method}d). In the following section, we summarize the parameters used in the supergranulation detection pipeline and discuss detection performance for individual maps under different levels of solar activity, including both highly active and quiet-Sun conditions.

\section{Summary of the Utilized Parameters for Identification of Supergranulation Patterns}\label{sec:performance}

Identifying supergranulation patterns relies on parameters that define detection spatial scales, thresholding criteria, and the filtering of non-supergranular features. Here we summarize the key parameters used in the analysis and their physical motivation.

In particular, identification of supergranules with the LoG algorithm was performed for a range of scales from 10 to 35 Mm to perform blob detection of structures that occupy the following size range in pixels in Eq.~\ref{eq:log1}, 
\begin{equation}
    \sigma_{\rm min,max}=\frac{d_{\rm min,max}}{2\sqrt{2}dx}.
\end{equation}
This range encompasses the characteristic sizes of supergranular cells at different stages of their evolution while excluding granulation and unrealistically large-scale flow structures. 

\begin{figure}[t]
\centerline{\includegraphics[width=0.95\textwidth,clip=]{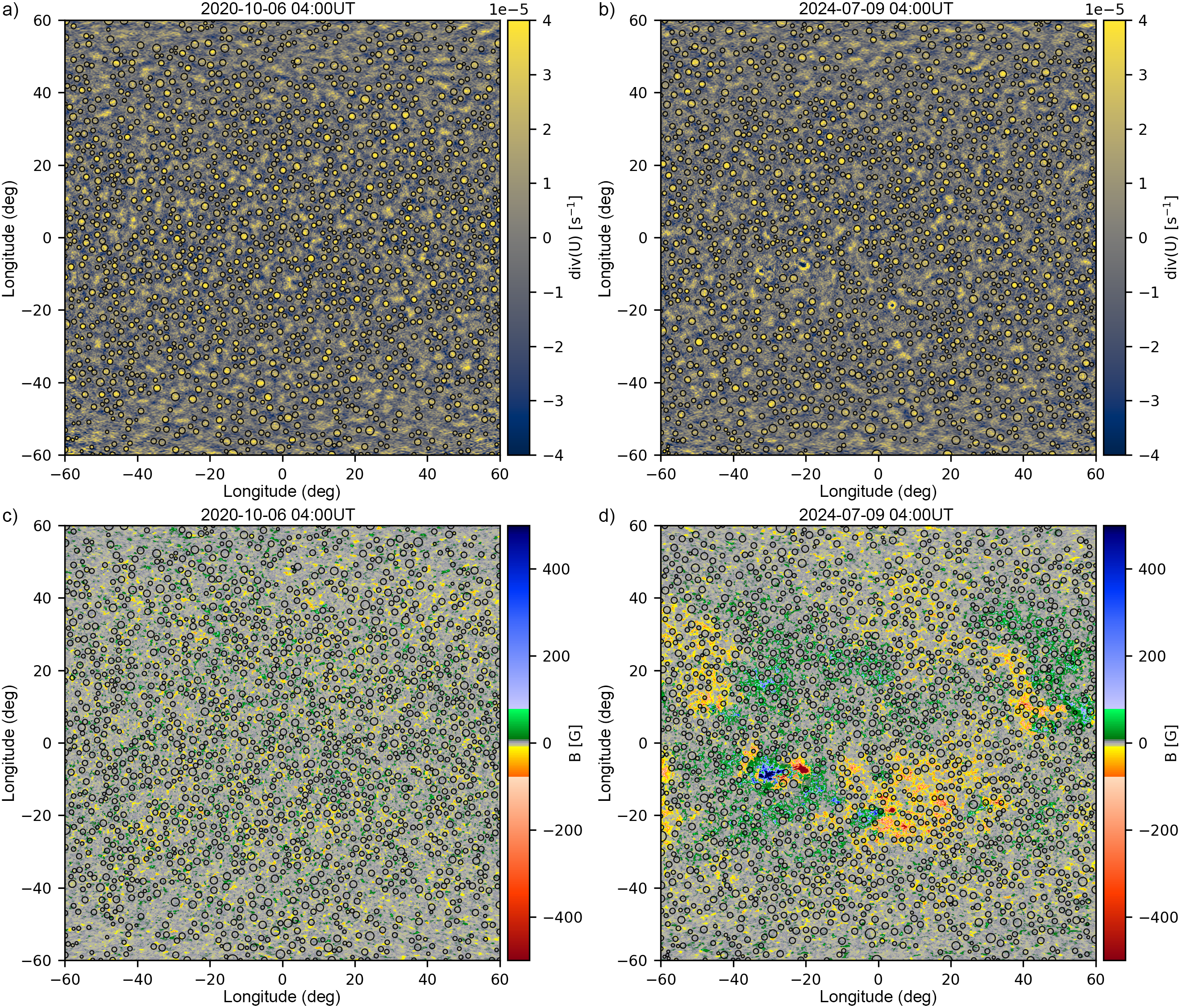}}
\caption{Example of supergranule identification for most extreme cases: the quiet Sun on 6 October, 2020 (panels a and c) and the highest magnetic activity over the SDO/HMI period of observations on 9 July, 2024 (panels b and d). Supergranules are indicated by circles over the velocity divergence maps (panels a and b) and line-of-sight magnetograms (panels c and d). The circle sizes correspond to the identified supergranules.}
\label{fig:minmax}
\end{figure}

The detection of candidate supergranular cells is performed using the LoG method applied to the normalized divergence field (Eq.~\ref{eq:log0}). The scale-space identification of structures was sampled with a step size of 0.25~Mm across the evaluated spatial scales, ensuring adequate coverage of supergranular sizes. The LoG response threshold is set to ``0", allowing all candidate structures to be detected at this stage. Overlapping detections were suppressed, ensuring that each structure is represented by a single, distinct detection.

To remove identifications associated with active regions, we used corresponding line-of-sight magnetograms to identify pixels with magnetic field strengths of $|B| \geq 1000$~G. To account for the spatial extent of active regions, associated moat flows, and the influence of strong-field regions, the resulting mask was expanded using a disk-shaped dilation kernel with a radius of 20~px ($\sim 28$~Mm). 

To ensure the physical relevance of the detected features, additional constraints were imposed based on horizontal flow divergence. In particular, only regions with positive divergence were considered, corresponding to outflow centers of supergranules. In addition, a noise-adaptive threshold was applied using $k = 3.0$ (Eq.~\ref{eq:thr}), requiring the divergence signal to exceed three times the characteristic background fluctuations. This approach provides a robust criterion for distinguishing coherent flow structures from noise. To further suppress spurious detections from fragmented or multi-peaked structures, a peak-consistency criterion was applied within the LoG-defined regions. Secondary local maxima were identified using a minimum spatial separation of 3 pixels. These peaks were required to exceed a 0.3 threshold of the primary peak amplitude. Structures exhibiting more than one significant peak were rejected, ensuring that each retained detection corresponds to a single, coherent supergranular cell.

Considering the use of velocity divergence maps, the applied spatial smoothing, and the intrinsic averaging introduced by the time-distance helioseismic inversion, the amplitudes and spatial extents of divergence features are reduced, leading to a systematic underestimation of supergranular radii (Eq.~\ref{eq:radius}). To compensate for this effect, a correction factor of $\alpha = 1.764$ is applied.

Testing the sensitivity of our results to variations in the detection and filtering parameters showed that changes in the LoG scale range, divergence threshold, multi-peak rejection criteria, active-region mask dilation, and size-correction factor did not affect the main conclusions of the study. The selection of parameters was tested for the two most extreme cases shown in Figure~\ref{fig:minmax}, corresponding to quiet-Sun and high-activity conditions, as well as for an intermediate level of solar activity. The results were evaluated both visually and through the statistical properties of the detected supergranule distributions.
In particular, selecting an adaptive 85th-percentile threshold for the divergence field, based on probability density functions evaluated over different time ranges (Figure~\ref{fig:hist_minmax}), allows us to eliminate statistically insignificant positive-divergence variations. Setting the LoG scale range too broad may detect very large structures ($\gtrsim 50$~Mm), whereas extending it toward smaller scales can capture background small-scale variations or structures associated with supergranule formation. Because the formation and evolution of supergranules are beyond the scope of this work, we do not attempt to determine whether a structure identified during its formation stage will subsequently evolve into a mature supergranule. The adopted lower limit of the LoG scale range nevertheless allows us to capture developing supergranules. To remove false-positive detections, we applied the multi-peak rejection criterion, based on the assumption that a developing supergranule should still appear as a distinct, coherent structure.

Following the detection procedure, we illustrate the performance of the algorithm under both quiet-Sun and high-activity conditions (Figure~\ref{fig:minmax}). For a very quiet Sun day (6 October 2020), the LoG algorithm initially identified 3250 candidate structures for the total available latitude range. The majority were rejected by the divergence thresholding, and an additional 300 candidates were removed by the multi-peak filtering. The final sample contains 1560 supergranules, which are overplotted on the divergence map and the corresponding magnetogram (Figure~\ref{fig:minmax}a,c). For a period of high solar activity (9 July 2024), the LoG algorithm detected 3486 structures. After applying the filtering steps, including removal of 1561 low-divergence candidates, 287 multi-peak structures, and 31 detections associated with active regions, the final sample consists of 1607 supergranules (Figure~\ref{fig:minmax}b,d). 

\section{Solar-Cycle Variability of Supergranulation}\label{sec:results}
\subsection{Temporal Evolution of Global Supergranular Properties}
\begin{figure}[b]
\centerline{\includegraphics[width=0.65\textwidth,clip=]{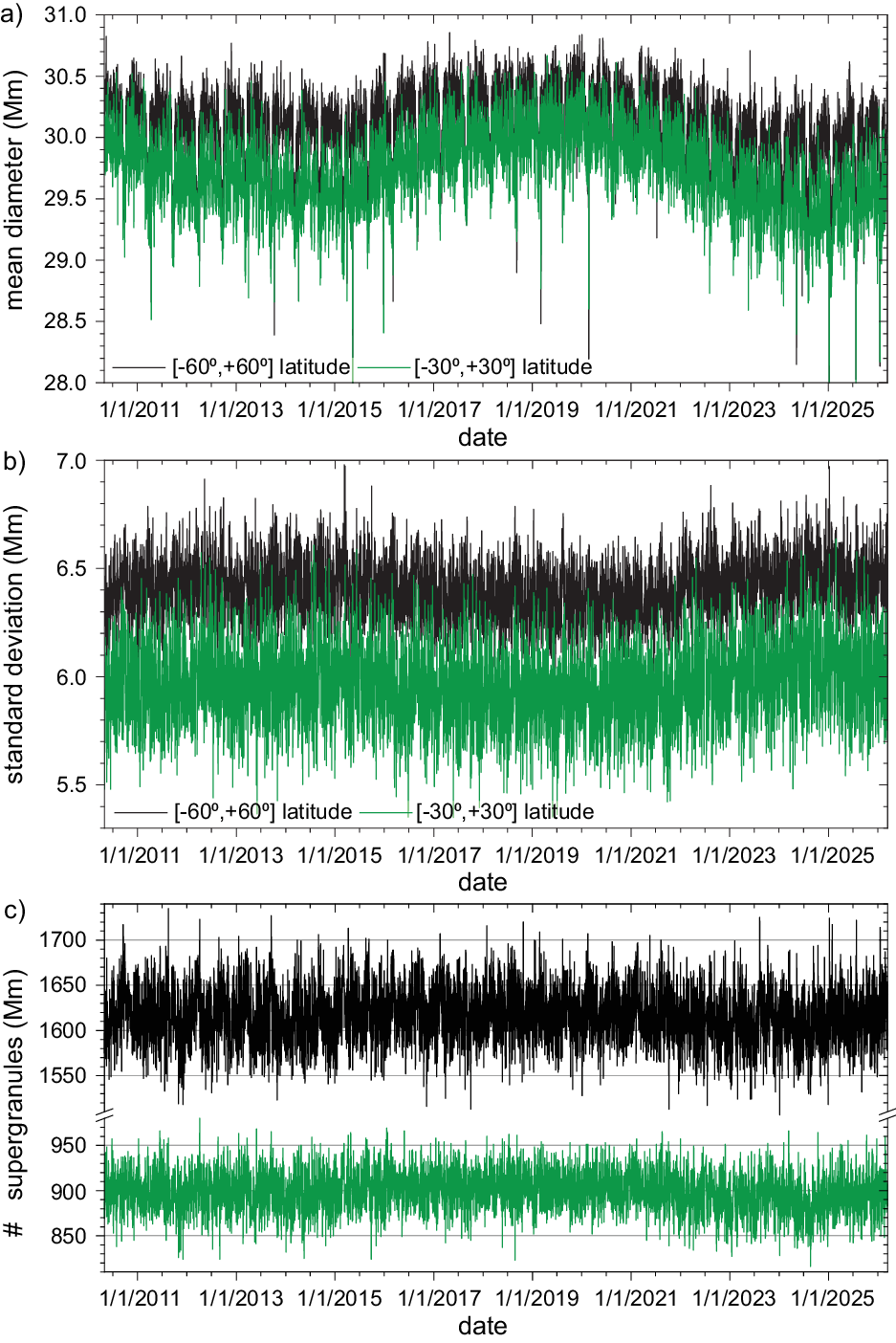}}
\caption{Temporal variations of the mean supergranular diameter (panel a), the standard deviation of the supergranular diameter distribution (panel b), and the number of identified supergranules (panel c) over Solar Cycles 24 and 25.}
\label{fig:Nsupergran-diam}
\end{figure}

\begin{figure}[t]
\centerline{\includegraphics[width=0.65\textwidth,clip=]{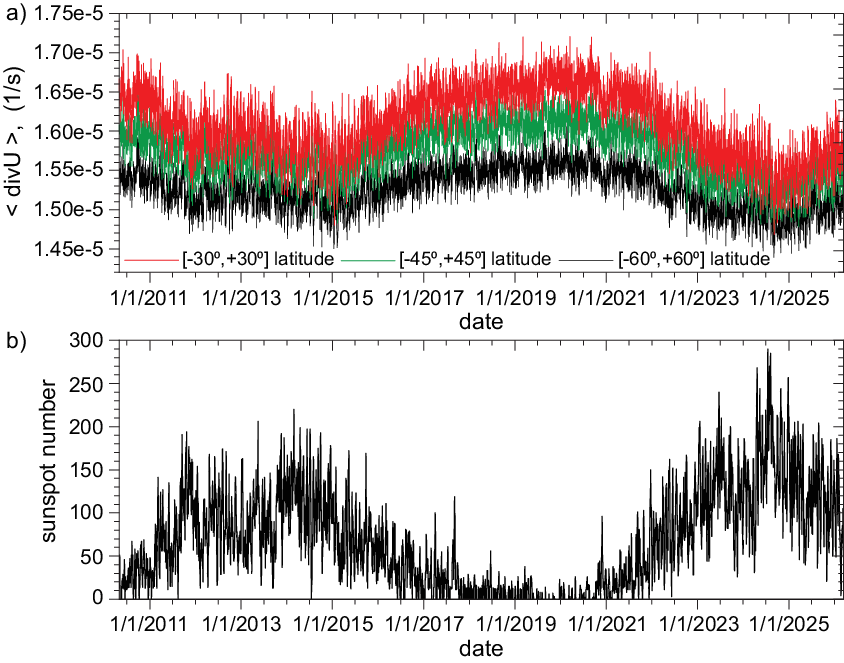}}
\caption{Temporal distribution of the mean velocity divergence for the identified supergranules (panel a) and the daily sunspot number (panel b) during solar cycles 24 and 25. Sunspot data from the World Data Center SILSO, Royal Observatory of Belgium, Brussels~\citep{SILSO_Sunspot_Number}.}
\label{fig:divU-SN}
\end{figure}

After validating the detection procedure under both quiet-Sun and active conditions, we applied the pipeline described in Section~\ref{sec:pipeline} to approximately 16 years of SDO/HMI observations spanning from 2010 May 1 to 2026 March 14. 
The presented analysis reveals a decrease in the mean supergranular diameter during the maxima of Solar Cycles 24 and 25, followed by an increase during periods of low magnetic activity (Figure~\ref{fig:Nsupergran-diam}a). This trend demonstrates that the detection procedure is sensitive enough to capture solar-cycle variations in supergranular properties, including differences in activity levels between the two cycles. In particular, Solar Cycle 24 was noticeably weaker than Solar Cycle 25, as reflected in the magnetic activity proxies (Figure~\ref{fig:divU-SN}a), and was associated with systematically larger mean supergranular diameters than those measured during the stronger Cycle 25.

Analysis of detections within different latitude ranges reveals a stronger solar-cycle dependence for latitudes within $\pm 30^{\mathrm{o}}$ (green curve; Figure~\ref{fig:Nsupergran-diam}a) than for the broader $\pm 60^{\mathrm{o}}$ range (black curve). This difference can be readily explained by the increasing contribution of regions outside the active-latitude belts, where active regions rarely form. Although ephemeral magnetic regions are also expected to suppress convective motions, their influence is likely weaker because their magnetic flux is spread over a larger area of the solar surface than the more compact magnetic concentrations associated with active regions. The decrease in the solar rotation rate with latitude may also contribute to the observed latitudinal variation in supergranular size. Together, these factors lead to a weaker solar-cycle dependence and a systematic increase in the mean supergranular diameter when a broader latitude range is considered.

A detailed investigation of the semiannual outliers in the supergranular size distribution (Figure~\ref{fig:Nsupergran-diam}a) indicates that these events are caused by increased noise in the input horizontal velocity maps rather than by physical changes in the characteristic supergranular scale. Although the velocity histograms and spatial power spectra show no pronounced differences, the affected maps exhibit substantially enhanced spatial gradients in both horizontal velocity components. This increases the noise level of the derived divergence field by about 20\% relative to neighboring days. Because supergranule identification is based on positive divergence regions, increased divergence noise produces an excess of small-scale detections near the lower diameter limit, artificially decreasing the mean detected supergranular size.

Foreshortening effects may also make supergranules appear larger toward higher latitudes. The LoG detection procedure partially mitigates this bias by assuming approximately circular structures. This assumption is consistent with the morphology of the detected supergranular cells in the divergence maps. Nevertheless, some fraction of the observed size increase may still be due to projection effects. 

The standard deviation of the supergranular diameter also exhibits systematic solar-cycle variations (Figure~\ref{fig:Nsupergran-diam}b), increasing during periods of high magnetic activity and decreasing around solar minima. This behavior indicates that the range of supergranular scales broadens during cycle maxima, likely reflecting more complex interactions between convective flows and magnetic fields. Furthermore, the standard deviation is systematically larger for the $\pm 60^{\mathrm{o}}$ latitude range (black curve) than for $\pm 30^{\mathrm{o}}$ (green curve), consistent with the inclusion of a wider variety of dynamical and magnetic environments. The larger latitudinal extent samples both active and quiet regions, as well as areas with different rotation rates and magnetic flux distributions, resulting in a broader distribution of detected supergranular sizes.

In addition to variations in the characteristic size of supergranules, the total number of detected cells exhibits a weak but noticeable dependence on the solar cycle (Figure~\ref{fig:Nsupergran-diam}c). The number of detections decreases slightly during periods of maximum magnetic activity and increases toward solar minima. Although the amplitude of this variation is significantly smaller than that observed for the mean diameter, the trend is consistent across both Solar Cycles 24 and 25. The relatively modest decrease in the number of detected supergranules suggests that magnetic activity primarily affects the supergranular flows rather than dramatically changing their overall occurrence rate. The reduction in detected cells may result from suppressed convective motions in magnetically active regions, where strong magnetic fields weaken or obscure the divergence signatures used to identify supergranules.

This interpretation is further supported by the temporal evolution of the velocity divergence associated with supergranular flows (Figure~\ref{fig:divU-SN}a). If magnetic fields suppress convective motions, their influence should be reflected not only in the number and size of detected supergranules, but also in the strength of the divergence signal itself. Indeed, considering the overall increase in magnetic flux during solar maximum, it is not surprising that the velocity divergence associated with supergranular flows decreases noticeably. The anticorrelation between the mean velocity divergence within detected supergranules and the sunspot number is substantially stronger than that found for the supergranule diameter. In general, an increase in the mean velocity divergence is well correlated with a decrease in the total sunspot number (Figure~\ref{fig:divU-SN}b).

The results presented in this section characterize the temporal evolution of supergranular properties averaged over selected latitude ranges. To build a more complete picture of their solar-cycle variability, we next examine how these properties vary with latitude.

\subsection{Latitudinal Dependence of Supergranular Properties}

The temporal trends discussed above represent averages over broad latitude ranges and therefore may obscure important latitudinal variations. To better resolve the relationship between supergranulation and magnetic activity, we examine time-latitude distributions of supergranular properties and flows, enabling direct comparison with the latitudinal migration of solar activity throughout Solar Cycles 24 and 25.

\begin{figure}[t]
\centerline{\includegraphics[width=1\textwidth,clip=]{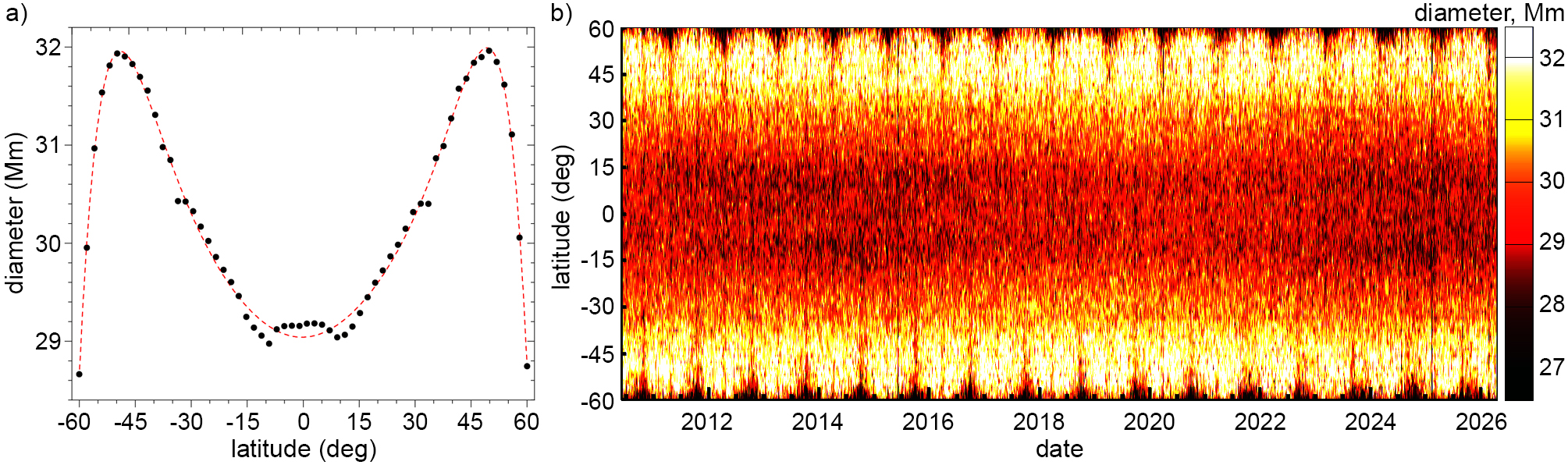}}
\caption{Latitudinal distribution of the mean supergranular diameter (panel a) and time–latitude map of the mean supergranular diameter (panel b) over Solar Cycles 24 and 25.}
\label{fig:lat-diam}
\end{figure}

In general, the mean diameter of supergranules averaged over the full period of SDO/HMI observations varies non-linearly with the latitude and is characterized by its gradual increase from the equator to higher latitudes and fast decrease in size above $50^{\mathrm{o}}$ (Figure~\ref{fig:lat-diam}a). The increase in diameter may reflect both the decrease in the solar rotation rate with latitude and the reduced influence of magnetic activity outside the active-latitude belts. Variations in the mean diameter near $\pm 9^{\mathrm{o}}$ and $\pm 34^{\mathrm{o}}$ are associated with the merging procedure of overlapping $15^{\mathrm{o}}$ tiles used to construct the $120^{\mathrm{o}} \times 120^{\mathrm{o}}$ maps \citep{Zhao2012}. The rapid decrease in measured supergranular size at latitudes above $\pm 50^{\mathrm{o}}$ likely results from foreshortening.

Such latitude-dependent variations are clearly visible in the time–latitude distribution of supergranular diameter (Figure~\ref{fig:lat-diam}b), which reveals the strongest size reduction within the solar activity belts during the maxima of Solar Cycles 24 and 25. In contrast, during solar minima, supergranules at low latitudes are systematically larger than during active phases of the solar cycle. At higher latitudes (around $\pm 50^{\mathrm{o}}$), a weak but still noticeable systematic reduction in supergranular size is also present. 

\begin{figure}[t]
\centerline{\includegraphics[width=0.7\textwidth,clip=]{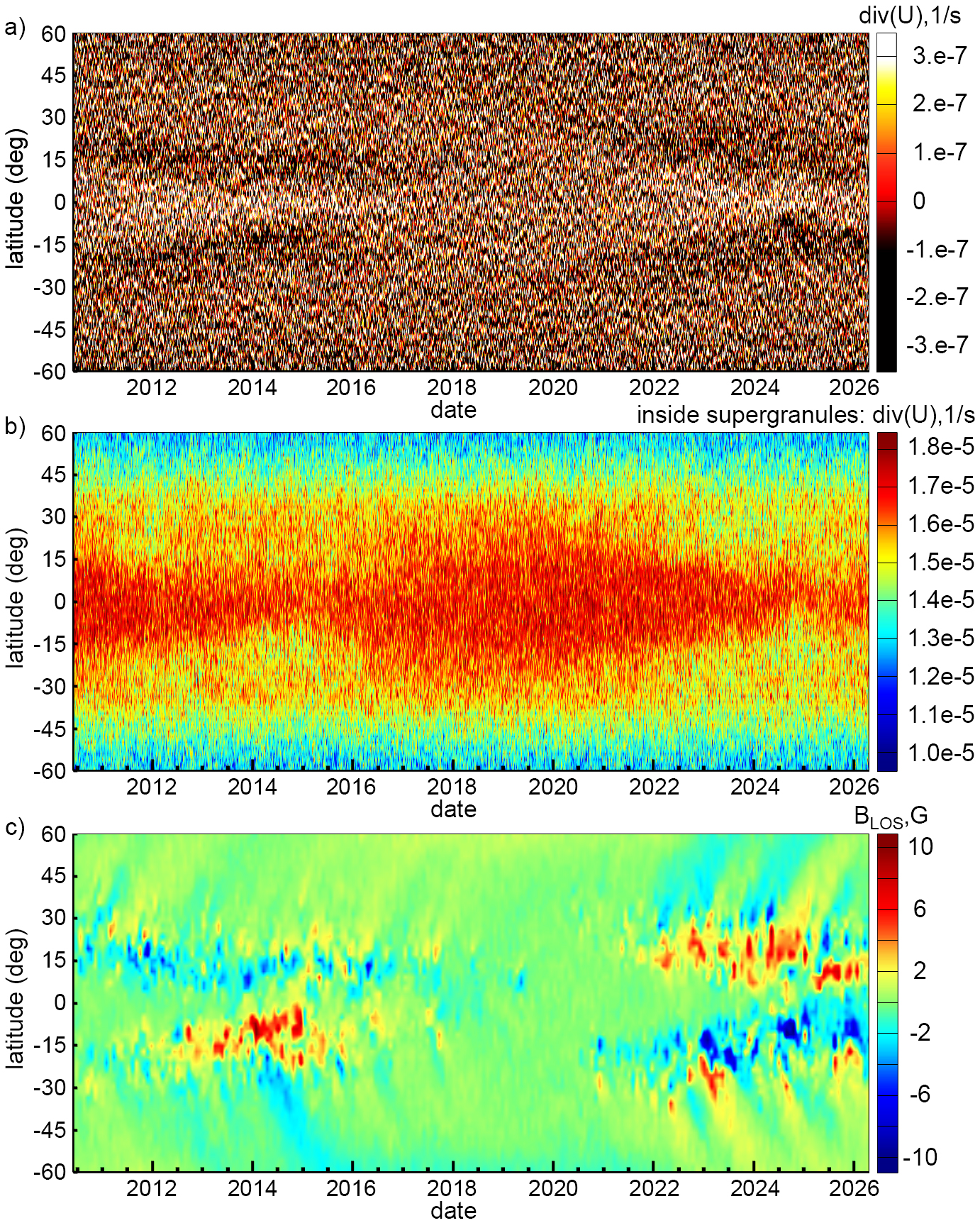}}
\caption{Time–latitude distributions of the mean horizontal velocity divergence calculated over the full field of view (panel a), the mean velocity divergence measured within detected supergranules (panel b), and the line-of-sight magnetic field (panel c).}
\label{fig:batterflyDivU}
\end{figure}

In addition to the solar-cycle variations, semiannual episodes of reduced supergranular size appear across nearly the entire latitude range. These events are unlikely to be associated with Earth’s orbital motion, as they occur primarily in January and July and do not coincide with periods of large $B_0$ angle. Their nearly uniform latitudinal distribution further suggests that they do not reflect genuine changes in solar convection or viewing-angle effects. Instead, detailed analysis indicates that these episodes arise from increased noise in the input horizontal velocity maps, which amplifies small-scale fluctuations in the derived divergence field and produces an excess of false detections near the lower diameter limit. This suggests that the observed semiannual decreases are most likely related to periodic calibration or processing effects in the HMI data products.

The time–latitude distribution of mean velocity divergence shows a butterfly-diagram pattern similar to that observed for the line-of-sight magnetic field pattern (Figure~\ref{fig:batterflyDivU}a,c), indicating suppressed convective flows in the bands of strong magnetic activity and a more homogeneous distribution outside these regions. In the case of the filtered velocity divergence within detected supergranules, a clear temporal and latitudinal dependence on magnetic activity is evident (Figure~\ref{fig:batterflyDivU}b). In general, supergranules with stronger divergence are concentrated at latitudes where magnetic activity is weakest, particularly in the regions between the activity branches (Figure~\ref{fig:batterflyDivU}c). In contrast, during solar minima, supergranules with stronger divergence occupy a much broader range of latitudes, reflecting the overall reduction of magnetic suppression of convective flows. A systematic decrease in the measured velocity divergence above $\pm 45^{\mathrm{o}}$ latitude is likely influenced by foreshortening effects and geometric distortions introduced by the Postel projection.

\begin{figure}[t]
\centerline{\includegraphics[width=1\textwidth,clip=]{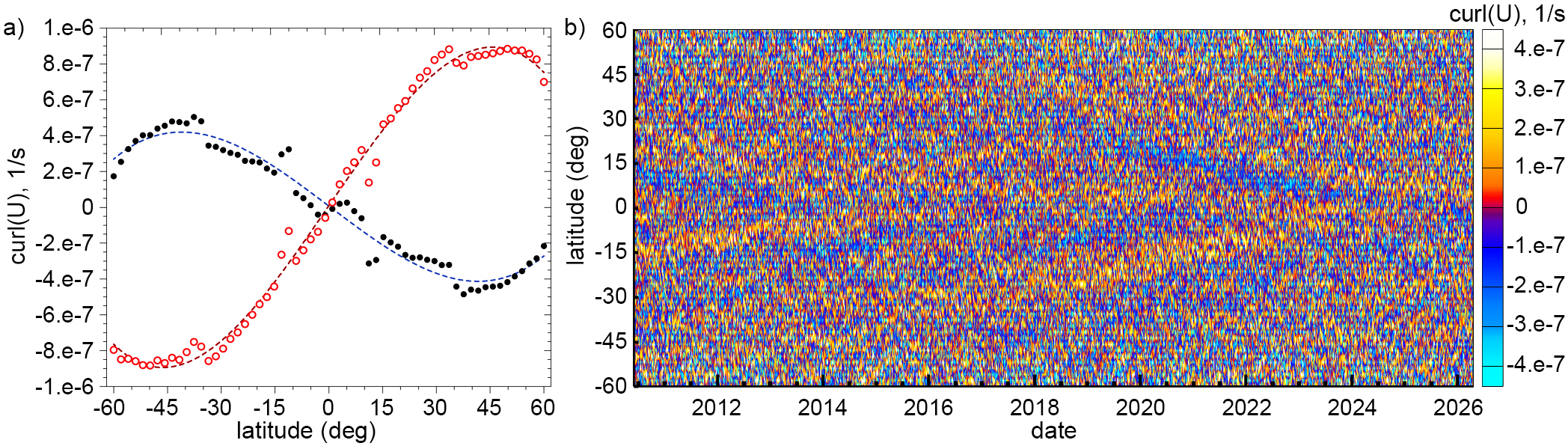}}
\caption{Latitudinal distributions of the mean curl of the horizontal velocity, curl(U), measured within identified supergranules (empty circles) and from the full field of view (black circles; panel a). Panel (b) shows the time–latitude map of the residual curl(U) obtained after subtracting the mean latitudinal profile shown in panel (a), over Solar Cycles 24 and 25.}
\label{fig:curlU}
\end{figure}

Given the complexity of interaction between multiscale flows, rotation, and magnetic fields, we also investigated the properties of the curl of the horizontal velocity, curl(U). In contrast to the clear correlation between velocity divergence and magnetic activity (Figure~\ref{fig:batterflyDivU}), no strong solar-cycle dependence is evident in curl(U). The filtered curl(U) measured within detected supergranules exhibits a pronounced hemispheric asymmetry, with predominantly negative values in the Southern hemisphere and positive values in the Northern hemisphere (empty circles; Figure~\ref{fig:curlU}a). Interestingly, the total curl(U) averaged over the full field of view shows the opposite hemispheric trend, but with significantly smaller amplitude (black circles). In the time-latitude distribution, solar-cycle-related variations become more apparent only after subtracting the mean latitudinal profile of full-field curl(U) (Figure~\ref{fig:curlU}b).

Taken together, the temporal and latitudinal analyses show that solar-cycle variability in supergranulation is strongly structured in both time and latitude. The global trends reveal systematic decreases in supergranular size and velocity divergence during periods of enhanced magnetic activity, while the latitude-resolved analysis shows that these variations are concentrated primarily within the active-latitude belts and closely follow the spatiotemporal migration of solar magnetic activity. In contrast, the vortical component of the flow exhibits much weaker solar-cycle dependence and is dominated by hemispheric asymmetry, suggesting a more complex interplay between convection, rotation, and magnetic fields. These results support the interpretation that supergranulation serves as a sensitive tracer of solar-cycle-dependent changes in near-surface convection.

\section{Discussion}\label{sec:discussion}
The interaction between magnetic fields and convective flows across a broad range of spatial scales drives many phenomena associated with solar variability. Supergranulation, as one of the dominant scales of near-surface convection, provides a valuable diagnostic for investigating this interaction over the solar activity cycle. In this work, we analyzed the temporal and latitudinal evolution of supergranular properties using approximately 16 years of SDO/HMI observations spanning Solar Cycles 24 and 25. To enable this analysis, we developed an automated detection pipeline based on Laplacian-of-Gaussian identification of localized positive-divergence regions in horizontal velocity-divergence maps, combined with magnetic-field masking and additional filtering to remove spurious detections and structures associated with active regions.

The resulting supergranule identifications show a clear correspondence with the magnetic network (Figure~\ref{fig:minmax}), a relationship widely used in previous observational studies of supergranulation (Table~\ref{tab:SG-obs}). However, it is important to note that supergranular sizes inferred from different diagnostics, including velocity divergence, feature tracking, intensity patterns, and magnetic-network elements, are not necessarily identical and may introduce systematic biases into the inferred size distribution. In particular, divergence-based measurements can depend strongly on the degree of smoothing, segmentation method, and detection thresholds. Therefore, we regard the characteristic supergranular scale of approximately 30~Mm, commonly inferred from analyses of turbulent spectra, as a robust reference value for comparison \citep[e.g.,][]{Rincon2018}. Within this observational and methodological context, our measurements are best interpreted in terms of relative temporal and latitudinal variations rather than as an absolute determination of the supergranular scale.

Our analysis reveals a clear solar-cycle dependence of supergranular size and flow properties. The mean diameter decreases systematically during solar maxima and increases during periods of low magnetic activity, indicating that enhanced magnetic activity suppresses the characteristic scale of supergranular convection (Figure~\ref{fig:Nsupergran-diam}a). Consistent with this trend, the weaker Solar Cycle 24 is associated with systematically larger supergranules than the stronger Solar Cycle 25. The temporal evolution of the diameter distribution further shows that the standard deviation increases during cycle maxima (Figure~\ref{fig:Nsupergran-diam}b), implying a broader range of detected supergranular scales under stronger magnetic activity. In contrast, the total number of detected supergranules exhibits only weak solar-cycle dependence (Figure~\ref{fig:Nsupergran-diam}c), suggesting that magnetic activity primarily modifies the properties of supergranular flows rather than their overall occurrence rate.

The mean latitudinal profile of supergranular diameter provides additional information about the physical and observational factors influencing the measured size distribution. The gradual increase in mean diameter from the equator toward mid-latitudes (Figure~\ref{fig:lat-diam}a) may be related to the decrease in solar rotation rate with latitude and to the reduced influence of strong magnetic activity outside the activity belts. At the same time, the rapid decrease in measured size above about $\pm 50^{\mathrm{o}}$ is likely affected by foreshortening and geometric distortions introduced by the Postel projection. Qualitatively similar latitudinal variations in supergranular size have been reported in previous studies \cite[e.g.,][]{Raju2002}. 

The latitude dependence of the solar-cycle signal is most clearly seen in the time-latitude distribution of supergranular diameter. The strongest reduction in size occurs within the active-latitude belts during the maxima of Solar Cycles 24 and 25 (Figure~\ref{fig:lat-diam}b). In contrast, at higher latitudes, where active regions are rare, the variations are weaker. This shows that the global solar-cycle variation in mean diameter is largely produced by localized suppression of supergranular-scale convection in magnetically active belts. The latitudinal dependence also helps explain why the solar-cycle signal weakens when broader latitude ranges are included: averaging over $\pm 60^{\mathrm{o}}$ mixes active belts with higher-latitude regions that experience substantially weaker magnetic modulation (Figure~\ref{fig:Nsupergran-diam}a).

The latitude dependence of the solar-cycle signal is most clearly seen in the time-latitude distribution of supergranular diameter Figure~\ref{fig:lat-diam}b). The systematic decrease in supergranular size toward lower latitudes is consistent with the general behavior of rotating convection, in which rotational constraint can suppress buoyancy-driven convective motions and reduce convective velocities \citep{Chandrasekhar1961}. In addition, the systematic reduction in size occurs near the active-latitude belts, suggesting that supergranulation is coupled to the large-scale evolution of the solar magnetic field, particularly in the low-latitude regions where the toroidal component of the dynamo-generated magnetic field emerges at the surface \citep{Parker1955,Babcock1961,Leighton1969}. However, considering that even during solar minimum the supergranular size remains reduced at latitudes corresponding to the activity belts relative to both the equator and higher latitudes (Figure~\ref{fig:lat-diam}b), while the velocity divergence does not show a comparable belt-like structure  (Figure~\ref{fig:batterflyDivU}b), the persistent latitudinal variation in supergranular size cannot be attributed solely to local surface magnetic activity. Instead, it may reflect an additional modulation associated with the large-scale solar dynamo. This effect becomes especially strong during the maxima of Solar Cycles 24 and 25, when substantial magnetic flux is present at the solar surface. At the same time, the reduction in supergranular size and horizontal velocity divergence correlates with bursts of magnetic activity (Figures~\ref{fig:lat-diam}b and~\ref{fig:batterflyDivU}b,c), indicating an additional influence of enhanced magnetic flux on supergranular convection at smaller spatial scales. Thus, the observed variations in supergranular size likely reflect the combined influence of hydrodynamic and magnetic effects operating from global to local spatial scales.

The strongest response to magnetic activity is observed in the divergent component of the horizontal flow. The mean velocity divergence within detected supergranules strongly anticorrelates with sunspot number (Figure~\ref{fig:divU-SN}), indicating that increasing magnetic flux suppresses divergent convective outflows. The time-latitude distribution of velocity divergence shows a butterfly-diagram pattern similar to that of the line-of-sight magnetic field (Figure~\ref{fig:batterflyDivU}), further supporting the interpretation that magnetic fields directly modulate supergranular outflows. 
Nevertheless, it is important to keep in mind that the amplitude and spatial structure of the inferred flows and the resulting divergence are affected by the finite spatial resolution of the time-distance helioseismic inversions, which can reduce the measured divergence amplitudes. Therefore, the absolute values of div(U) should be interpreted with caution. In the presented analysis, the same inversion and processing procedures are applied consistently throughout the entire data set; therefore, the relative temporal and latitudinal variations are expected to be considerably less sensitive to these effects.
This result is particularly important because the divergence signal measures the flow field itself rather than the magnetic network accumulated by the flow. Therefore, the close correspondence between div(U) and the magnetic butterfly diagram provides direct evidence that the near-surface convective flow field responds to the evolving distribution of solar magnetic activity.

In contrast to the divergent component, the curl of the horizontal velocity exhibits substantially weaker solar-cycle dependence. Rather than following solar-cycle modulation, the filtered curl(U) within detected supergranules is primarily characterized by a pronounced hemispheric asymmetry, with predominantly negative values in the Southern hemisphere and positive values in the Northern hemisphere (Figure~\ref{fig:curlU}a). The full-field curl(U) shows the opposite hemispheric trend, but with significantly smaller amplitude. This difference between divergence and vorticity suggests that magnetic activity affects supergranular outflows more directly than rotational components of the flow. The hemispheric organization of curl(U) instead points to the influence of rotation, Coriolis forces, and large-scale flow organization. Thus, the divergent and vortical components appear to encode different aspects of the coupling between convection, rotation, and magnetic fields.

Several observational and methodological limitations should be considered when interpreting these results. Semiannual episodes of reduced supergranular size appear across nearly the entire latitude range and are likely associated with periodic increases in noise in the horizontal velocity maps rather than physical changes in convection (Figure~\ref{fig:Nsupergran-diam}a and Figure~\ref{fig:lat-diam}b). In addition, foreshortening and projection effects likely influence the measured sizes and divergence amplitudes at high latitudes, particularly above $\pm 45^{\mathrm{o}}$--$\pm 50^{\mathrm{o}}$. Although the LoG detection procedure partially mitigates these effects by identifying approximately circular divergence structures, a more quantitative assessment will require comparison with realistic numerical simulations and observations obtained from off-ecliptic viewpoints, such as those provided by Solar Orbiter \citep{Muller2013}. Future work should also examine the depth dependence of these trends, since supergranular-scale flows inferred at different depths may respond differently to magnetic activity and rotation.

The observed decrease in supergranular diameter during periods of enhanced activity is consistent with studies that reported an anticorrelation between supergranular size and magnetic activity using Ca~II~K images, Doppler measurements, or flow-derived diagnostics (Table~\ref{tab:SG-obs}). However, other studies have reported positive correlations or no clear relationship. For example, using horizontal flows derived from granule tracking in selected quiet-Sun regions near the disk center, \citet{Roudier2017} found no significant variation in meso- and supergranular flow properties over Solar Cycle 24. This result is consistent with our finding that the strongest solar-cycle modulation is concentrated within the active-latitude belts, while regions outside these belts exhibit substantially weaker variations. A similar point is supported by Ca~II~K measurements \citep{Raju2002}, which show that the characteristic supergranular scale depends on latitude even during relatively low-activity conditions. In our analysis, this latitude dependence is also evident in the time-latitude distribution of supergranular diameter (Figure~\ref{fig:lat-diam}b).

The opposite trends reported in some previous studies may partly reflect differences in the observables and detection methods used to define supergranular cells. In particular, studies based on magnetic-network or chromospheric-emission tracers may measure the organization of magnetic flux at cell boundaries rather than the convective-flow pattern itself. For example, \citet{Muenzer1989} found that Ca~II~K network-cell sizes increase with local chromospheric activity, indicating that magnetic-network tracers can respond differently from divergence-based flow diagnostics. Watershed-based methods may introduce an additional bias because they divide the observed field into complete cellular structures using local extrema and cell boundaries. If magnetic activity weakens small-scale flow signatures, modifies the magnetic network, or makes the boundaries between neighboring cells less distinct, several adjacent cells may be grouped into a larger structure. In this case, the measured cell size can increase with activity even if the underlying divergent-flow pattern is suppressed or fragmented. A similar effect may influence autocorrelation-based analyses \citep[e.g.,][]{Sykora1970}, where reduced contrast or merging of network features can shift the inferred characteristic scale toward larger values. Because autocorrelation methods measure a dominant spatial scale rather than identify individual supergranular cells, they have limited ability to determine whether a broad correlation peak corresponds to one large structure or to multiple unresolved neighboring cells. These methodological differences provide a plausible explanation for the diversity of reported correlations. In contrast to magnetic-network and chromospheric diagnostics, the present work identifies supergranules through the dynamical signature of positive horizontal velocity divergence in subsurface flow maps. Thus, our results suggest that magnetic activity suppresses the convective-flow signature of supergranulation, even though magnetic-network tracers may respond differently.

It is important to note that the reported size of supergranules depends on the observational diagnostic and analysis method. For example, a gradient-based analysis of SOHO/MDI intensity data found characteristic sizes of about 17.5~Mm during low solar activity and about 15.5~Mm during periods of high sunspot number \citep{Meunier2008}. Measurements based on local correlation tracking and velocity-divergence maps have reported characteristic scales in the range of 12-20~Mm \citep[e.g.,][]{DeRosa2002}, whereas values closer to 30~Mm have been obtained from other analyses of horizontal flows \citep{Meunier2007c} and from turbulent power spectra \citep{Rieutord2010a}. These differences demonstrate that the inferred scale is sensitive to smoothing, segmentation criteria, and the specific definition of a supergranular cell \citep{Meunier2007c}.

In the case of velocity divergence derived from time-distance helioseismology, the identified structures can be smaller than the commonly quoted $\sim 30$~Mm scale because the detection is based on positive-divergence cores and because the flow maps are affected by spatial smoothing and intrinsic averaging associated with the helioseismic inversions \citep[e.g.,][]{Zhao2012,Rincon2018}. In our analysis, the most common uncorrected supergranular sizes inferred from the SDO/HMI helioseismic flow maps range from 16.4 to 17.1~Mm. Therefore, following previous studies of solar turbulent flows \citep{Rieutord2010a} and the general convergence toward a characteristic scale of about 30~Mm, we apply a correction factor of $\alpha = 1.764$ to calibrate the LoG-derived size estimates to this reference scale.

With this methodological context in mind, the main conclusions of the analysis are based primarily on the relative temporal and latitudinal evolution of the detected structures rather than on the absolute calibrated size alone. Overall, the results indicate that supergranulation is not merely a passive convective pattern, but a sensitive dynamical tracer of solar-cycle variability. The characteristic size and divergent component of supergranular flows respond strongly to the evolving distribution of magnetic activity, with the strongest modulation concentrated in the active-latitude belts. In contrast, the vortical component exhibits weaker solar-cycle dependence and stronger hemispheric organization, suggesting that additional dynamical processes shape supergranular vorticity.

\section{Conclusions}\label{sec:conclusion}
We have investigated the solar-cycle variability of supergranulation using approximately 16 years of SDO/HMI time-distance helioseismology observations spanning Solar Cycles 24 and 25. For this purpose, we developed an automated detection pipeline based on Laplacian-of-Gaussian identification of localized positive-divergence regions in horizontal velocity-divergence maps, combined with magnetic-field masking, divergence thresholding, and rejection of multi-peaked structures. This approach enables a consistent characterization of supergranular properties under both quiet-Sun and active conditions. Applying this methodology over the full observational interval reveals systematic solar-cycle variations in both the properties of individual supergranules and the underlying flow field, indicating a close connection between supergranular convection and magnetic activity.

Analysis of flow properties shows that horizontal velocity divergence provides the clearest signature of this connection. The mean divergence within detected supergranules is strongly anticorrelated with sunspot number, and its time-latitude distribution exhibits a butterfly-diagram-like pattern similar to that of the line-of-sight magnetic field. In contrast, curl(U) shows much weaker solar-cycle dependence and is dominated by hemispheric asymmetry, indicating that the divergent and vortical components of supergranular flows respond differently to magnetic activity, rotation, and large-scale flow organization.

In addition to these variations in the flow field, the analysis shows that the mean supergranular diameter varies systematically over the solar cycle. Supergranules are smaller during periods of enhanced magnetic activity and larger during solar minima, with systematically larger mean diameters during the weaker Solar Cycle 24 than during the stronger Solar Cycle 25. The width of the size distribution also increases during cycle maxima, whereas the total number of detected supergranules shows only weak solar-cycle dependence. These results indicate that magnetic activity primarily modifies the characteristic size and flow properties of supergranulation, rather than strongly changing their occurrence rate.

The results of our analysis suggest that several factors may influence the characteristic size of supergranules across different spatial and temporal scales. On global scales, rotation and large-scale magnetic-field organization may contribute to the systematic decrease in supergranular size toward lower latitudes. This large-scale latitudinal dependence may reflect rotationally constrained convection and the latitude-dependent organization of the global magnetic field, both of which can modify the characteristic scales and dynamics of near-surface convective flows. The persistence of this latitudinal pattern even during periods of low magnetic activity suggests that it cannot be explained solely by local surface magnetic fields and may also reflect large-scale dynamical or global dynamo-related modulation. On smaller scales, particularly within spatially extended magnetically active regions, such as active regions and plages, enhanced magnetic flux plays a significant role in reducing supergranular size and shaping the observed solar-cycle-dependent variations. 
This effect is most evident within the active-latitude belts, where the strongest reduction in supergranular size coincides with enhanced magnetic activity and follows its equatorward migration over the solar cycle. At higher latitudes, where strong magnetic activity is less frequent, the variations in supergranular size are substantially weaker. These results therefore suggest that the observed evolution of supergranular scales reflects a combination of large-scale dynamical effects and localized magnetic suppression of convection, with their relative contributions varying with latitude and solar-cycle phase.

Overall, the presented results demonstrate that supergranulation is strongly coupled to solar magnetic activity on multiple spatial and temporal scales and can serve as a sensitive tracer of solar-cycle-dependent changes in near-surface convection. The characteristic size and divergent component of supergranular flows respond directly to the evolving distribution of magnetic activity, while the vortical component reflects additional dynamical influences. Future work combining long-term helioseismic observations, realistic radiative MHD simulations, and off-ecliptic observations will help quantify projection effects, explore the depth dependence of these trends, and clarify the physical mechanisms governing the evolution of supergranulation over the solar cycle.

\begin{acknowledgments}
Resources supporting this work were provided by the NASA High-End Computing (HEC) Program through the NASA Advanced Supercomputing (NAS) Division at Ames Research Center. 
This investigation was supported by the NASA Heliophysics Guest Investigator Open Program (23-HGIO23\_2-0077) and the NASA Science DRIVE Center Program (COFFIES Project; 80NSSC22M0162). The data used in this work were obtained from the Joint Science Operations Center (JSOC) Science Data Processing pipeline. The Helioseismic and Magnetic Imager observations are courtesy of NASA/SDO and the HMI science team. Sunspot number data were obtained from the World Data Center SILSO, Royal Observatory of Belgium, Brussels. The authors also thank Junwei Zhao for providing the merged dataset derived from the SDO/HMI time-distance helioseismology data products. 
\end{acknowledgments}

%\bibliography{supergranulation}{}

\begin{thebibliography}{}
	\expandafter\ifx\csname natexlab\endcsname\relax\def\natexlab#1{#1}\fi
	\providecommand{\url}[1]{\href{#1}{#1}}
	\providecommand{\dodoi}[1]{doi:~\href{http://doi.org/#1}{\nolinkurl{#1}}}
	\providecommand{\doeprint}[1]{\href{http://ascl.net/#1}{\nolinkurl{http://ascl.net/#1}}}
	\providecommand{\doarXiv}[1]{\href{https://arxiv.org/abs/#1}{\nolinkurl{https://arxiv.org/abs/#1}}}
	
	% type= article
	\bibitem[{H.~W. {Babcock}(1961){Babcock}}]{Babcock1961}
	{Babcock}, H.~W. 1961, \bibinfo{title}{{The Topology of the Sun's Magnetic
			Field and the 22-Year Cycle.},} \apj, 133, 572, \dodoi{10.1086/147060}
	
	% type= article
	\bibitem[{J.~G. {Beck} \& J. {Schou}(2000){Beck} \& {Schou}}]{Beck2000}
	{Beck}, J.~G., \& {Schou}, J. 2000, \bibinfo{title}{{Supergranulation
			rotation},} \solphys, 193, 333, \dodoi{10.1023/A:1005258123855}
	
	% type= article
	\bibitem[{F. {Berrilli} {et~al.}(1999){Berrilli}, {Ermolli}, {Florio}, \&
		{Pietropaolo}}]{Berrilli1999}
	{Berrilli}, F., {Ermolli}, I., {Florio}, A., \& {Pietropaolo}, E. 1999,
	\bibinfo{title}{{Average properties and temporal variations of the geometry
			of solar network cells},} \aap, 344, 965
	
	% type= book
	\bibitem[{S. Chandrasekhar(1961)Chandrasekhar}]{Chandrasekhar1961}
	Chandrasekhar, S. 1961, Hydrodynamic and Hydromagnetic Stability (Oxford:
	Clarendon Press)
	
	% type= article
	\bibitem[{S. {Chatterjee} {et~al.}(2017){Chatterjee}, {Mandal}, \&
		{Banerjee}}]{Chatterjee2017}
	{Chatterjee}, S., {Mandal}, S., \& {Banerjee}, D. 2017,
	\bibinfo{title}{{Variation of Supergranule Parameters with Solar Cycles:
			Results from Century-long Kodaikanal Digitized Ca II K Data},} \apj, 841, 70,
	\dodoi{10.3847/1538-4357/aa709d}
	
	% type= misc
	\bibitem[{F. {Clette} \& L. {Lefèvre}(2015){Clette} \&
		{Lefèvre}}]{SILSO_Sunspot_Number}
	{Clette}, F., \& {Lefèvre}, L. 2015, SILSO Sunspot Number V2.0,,
	https://doi.org/10.24414/qnza-ac80
	
	% type= article
	\bibitem[{M.~L. {De Rosa} {et~al.}(2002){De Rosa}, {Gilman}, \&
		{Toomre}}]{DeRosa2002}
	{De Rosa}, M.~L., {Gilman}, P.~A., \& {Toomre}, J. 2002, \bibinfo{title}{{Solar
			Multiscale Convection and Rotation Gradients Studied in Shallow Spherical
			Shells},} \apj, 581, 1356, \dodoi{10.1086/344295}
	
	% type= article
	\bibitem[{T.~L. {Duvall}(1980){Duvall}}]{Duvall1980}
	{Duvall}, Jr., T.~L. 1980, \bibinfo{title}{{The Equatorial Rotation Rate of the
			Supergranulation Cells},} \solphys, 66, 213, \dodoi{10.1007/BF00150578}
	
	% type= article
	\bibitem[{H.~J. {Hagenaar} {et~al.}(1997){Hagenaar}, {Schrijver}, \&
		{Title}}]{Hagenaar1997}
	{Hagenaar}, H.~J., {Schrijver}, C.~J., \& {Title}, A.~M. 1997,
	\bibinfo{title}{{The Distribution of Cell Sizes of the Solar Chromospheric
			Network},} \apj, 481, 988, \dodoi{10.1086/304066}
	
	% type= article
	\bibitem[{C. {Huang} {et~al.}(2012){Huang}, {Yan}, {Zhang}, {Tan}, \&
		{Li}}]{Huang2012}
	{Huang}, C., {Yan}, Y., {Zhang}, Y., {Tan}, B., \& {Li}, G. 2012,
	\bibinfo{title}{{The Morphologic Properties of Magnetic Networks over the
			Solar Cycle 23},} \apj, 759, 106, \dodoi{10.1088/0004-637X/759/2/106}
	
	% type= inproceedings
	\bibitem[{S.~M. {Jefferies} {et~al.}(1988){Jefferies}, {Pomerantz}, {Duvall},
		{Harvey}, \& {Jaksha}}]{Jefferies1988}
	{Jefferies}, S.~M., {Pomerantz}, M.~A., {Duvall}, Jr., T.~L., {Harvey}, J.~W.,
	\& {Jaksha}, D.~B. 1988, \bibinfo{title}{{Helioseismology from the South
			Pole: comparison of 1987 and 1981 results.},} in ESA Special Publication,
	Vol. 286, Seismology of the Sun and Sun-Like Stars, ed. E.~J. {Rolfe},
	279--284
	
	% type= article
	\bibitem[{R. {Kariyappa} \& K.~R. {Sivaraman}(1994){Kariyappa} \&
		{Sivaraman}}]{Kariyappa1994}
	{Kariyappa}, R., \& {Sivaraman}, K.~R. 1994, \bibinfo{title}{{Variability of
			the Solar Chromospheric Network Over the Solar Cycle},} \solphys, 152, 139,
	\dodoi{10.1007/BF01473196}
	
	% type= article
	\bibitem[{Y. {Kimori}(2013){Kimori}}]{Kimori2013}
	{Kimori}, Y. 2013, \bibinfo{title}{{Morphological image processing for
			quantitative shape analysis of biomedical structures: effective contrast
			enhancement},} Journal of Synchrotron Radiation, 20, 848,
	\dodoi{10.1107/S0909049513020761}
	
	% type= article
	\bibitem[{R.~W. {Komm} {et~al.}(1995){Komm}, {Howard}, \& {Harvey}}]{Komm1995}
	{Komm}, R.~W., {Howard}, R.~F., \& {Harvey}, J.~W. 1995,
	\bibinfo{title}{{Characteristic Size and Diffusion of Quiet Sun Magnetic
			Patterns},} \solphys, 158, 213, \dodoi{10.1007/BF00795658}
	
	% type= article
	\bibitem[{R.~B. {Leighton}(1969){Leighton}}]{Leighton1969}
	{Leighton}, R.~B. 1969, \bibinfo{title}{{A Magneto-Kinematic Model of the Solar
			Cycle},} \apj, 156, 1, \dodoi{10.1086/149943}
	
	% type= article
	\bibitem[{R.~B. Leighton {et~al.}(1962)Leighton, Noyes, \&
		Simon}]{Leighton1962}
	Leighton, R.~B., Noyes, R.~W., \& Simon, G.~W. 1962, \bibinfo{title}{Velocity
		Fields in the Solar Atmosphere. I. Preliminary Report.,} The Astrophysical
	Journal, 135, 474, \dodoi{10.1086/147285}
	
	% type= book
	\bibitem[{T. Lindeberg(1994)Lindeberg}]{Lindeberg1994}
	Lindeberg, T. 1994, Scale-Space Theory in Computer Vision (Dordrecht: Kluwer
	Academic Publishers)
	
	% type= article
	\bibitem[{T. {Lindeberg}(1998){Lindeberg}}]{Lindeberg1998}
	{Lindeberg}, T. 1998, \bibinfo{title}{{Feature detection with automatic scale
			selection},} International Journal of Computer Vision, 30, 79–116,
	\dodoi{10.1023/A:1008045108935}
	
	% type= article
	\bibitem[{S. {Mandal} {et~al.}(2017){Mandal}, {Chatterjee}, \&
		{Banerjee}}]{Mandal2017}
	{Mandal}, S., {Chatterjee}, S., \& {Banerjee}, D. 2017,
	\bibinfo{title}{{Association of Supergranule Mean Scales with Solar Cycle
			Strengths and Total Solar Irradiance},} \apj, 844, 24,
	\dodoi{10.3847/1538-4357/aa76e3}
	
	% type= article
	\bibitem[{D. Marr \& E. Hildreth(1980)Marr \& Hildreth}]{Marr1980}
	Marr, D., \& Hildreth, E. 1980, \bibinfo{title}{Theory of edge detection,}
	Proceedings of the Royal Society of London. Series B. Biological Sciences,
	207, 187, \dodoi{10.1098/rspb.1980.0020}
	
	% type= article
	\bibitem[{S.~W. {McIntosh} {et~al.}(2011){McIntosh}, {Leamon}, {Hock}, {Rast},
		\& {Ulrich}}]{McIntosh2011}
	{McIntosh}, S.~W., {Leamon}, R.~J., {Hock}, R.~A., {Rast}, M.~P., \& {Ulrich},
	R.~K. 2011, \bibinfo{title}{{Observing Evolution in the Supergranular Network
			Length Scale During Periods of Low Solar Activity},} \apjl, 730, L3,
	\dodoi{10.1088/2041-8205/730/1/L3}
	
	% type= article
	\bibitem[{N. {Meunier}(2003){Meunier}}]{Meunier2003}
	{Meunier}, N. 2003, \bibinfo{title}{{Statistical properties of magnetic
			structures: Their dependence on scale and solar activity},} \aap, 405, 1107,
	\dodoi{10.1051/0004-6361:20030713}
	
	% type= article
	\bibitem[{N. {Meunier} \& A.-M. {Lagrange}(2020){Meunier} \&
		{Lagrange}}]{Meunier2020}
	{Meunier}, N., \& {Lagrange}, A.-M. 2020, \bibinfo{title}{{The effects of
			granulation and supergranulation on Earth-mass planet detectability in the
			habitable zone around F6-K4 stars},} \aap, 642, A157,
	\dodoi{10.1051/0004-6361/202038376}
	
	% type= article
	\bibitem[{N. {Meunier} {et~al.}(2008){Meunier}, {Roudier}, \&
		{Rieutord}}]{Meunier2008}
	{Meunier}, N., {Roudier}, T., \& {Rieutord}, M. 2008,
	\bibinfo{title}{{Supergranules over the solar cycle},} \aap, 488, 1109,
	\dodoi{10.1051/0004-6361:20078835}
	
	% type= article
	\bibitem[{N. {Meunier} {et~al.}(2007){Meunier}, {Tkaczuk}, {Roudier}, \&
		{Rieutord}}]{Meunier2007c}
	{Meunier}, N., {Tkaczuk}, R., {Roudier}, T., \& {Rieutord}, M. 2007,
	\bibinfo{title}{{Velocities and divergences as a function of supergranule
			size},} \aap, 461, 1141, \dodoi{10.1051/0004-6361:20065625}
	
	% type= article
	\bibitem[{H. {Muenzer} {et~al.}(1989){Muenzer}, {Schroeter}, {Woehl}, \&
		{Hanslmeier}}]{Muenzer1989}
	{Muenzer}, H., {Schroeter}, E.~H., {Woehl}, H., \& {Hanslmeier}, A. 1989,
	\bibinfo{title}{{Pole-equator-difference of the size of the chromospheric CA
			II-K-network in quiet and active solar regions},} \aap, 213, 431
	
	% type= article
	\bibitem[{D. {M{\"u}ller} {et~al.}(2013){M{\"u}ller}, {Marsden}, {St. Cyr},
		{Gilbert}, \& {Solar Orbiter Team}}]{Muller2013}
	{M{\"u}ller}, D., {Marsden}, R.~G., {St. Cyr}, O.~C., {Gilbert}, H.~R., \&
	{Solar Orbiter Team}. 2013, \bibinfo{title}{{Solar Orbiter. Exploring the
			Sun-Heliosphere Connection},} \solphys, 285, 25,
	\dodoi{10.1007/s11207-012-0085-7}
	
	% type= article
	\bibitem[{N.~K. {O'Sullivan} {et~al.}(2025){O'Sullivan}, {Aigrain},
		{Cretignier}, {Lakeland}, {Klein}, {Dumusque}, {Meunier}, {Sulis}, {Bedell},
		{Mortier}, {Cameron}, \& {Cegla}}]{OSullivan2025}
	{O'Sullivan}, N.~K., {Aigrain}, S., {Cretignier}, M., {et~al.} 2025,
	\bibinfo{title}{{Measuring the Sun's radial velocity variability due to
			supergranulation over a magnetic cycle.},} \mnras, 541, 3942,
	\dodoi{10.1093/mnras/staf1168}
	
	% type= article
	\bibitem[{E.~N. {Parker}(1955){Parker}}]{Parker1955}
	{Parker}, E.~N. 1955, \bibinfo{title}{{Hydromagnetic Dynamo Models.},} \apj,
	122, 293, \dodoi{10.1086/146087}
	
	% type= article
	\bibitem[{G. {Rajani} {et~al.}(2022){Rajani}, {Sowmya}, {Paniveni}, \&
		{Srikanth}}]{Rajani2022}
	{Rajani}, G., {Sowmya}, G.~M., {Paniveni}, U., \& {Srikanth}, R. 2022,
	\bibinfo{title}{{Solar Supergranular Fractal Dimension Dependence on the
			Solar Cycle Phase},} Research in Astronomy and Astrophysics, 22, 045006,
	\dodoi{10.1088/1674-4527/ac5020}
	
	% type= article
	\bibitem[{K.~P. {Raju} \& J. {Singh}(2002){Raju} \& {Singh}}]{Raju2002}
	{Raju}, K.~P., \& {Singh}, J. 2002, \bibinfo{title}{{Dependence of
			Supergranular Length-Scales on Network Magnetic Fields},} \solphys, 207, 11,
	\dodoi{10.1023/A:1015585010078}
	
	% type= article
	\bibitem[{M. {Rieutord} {et~al.}(2010){Rieutord}, {Roudier}, {Rincon},
		{Malherbe}, {Meunier}, {Berger}, \& {Frank}}]{Rieutord2010a}
	{Rieutord}, M., {Roudier}, T., {Rincon}, F., {et~al.} 2010, \bibinfo{title}{{On
			the power spectrum of solar surface flows},} \aap, 512, A4,
	\dodoi{10.1051/0004-6361/200913303}
	
	% type= article
	\bibitem[{F. {Rincon} \& M. {Rieutord}(2018){Rincon} \&
		{Rieutord}}]{Rincon2018}
	{Rincon}, F., \& {Rieutord}, M. 2018, \bibinfo{title}{{The Sun's
			supergranulation},} Living Reviews in Solar Physics, 15, 6,
	\dodoi{10.1007/s41116-018-0013-5}
	
	% type= article
	\bibitem[{T. {Roudier} {et~al.}(2017){Roudier}, {Malherbe}, \&
		{Mirouh}}]{Roudier2017}
	{Roudier}, T., {Malherbe}, J.~M., \& {Mirouh}, G.~M. 2017,
	\bibinfo{title}{{Dynamics of the photosphere along the solar cycle from
			SDO/HMI},} \aap, 598, A99, \dodoi{10.1051/0004-6361/201629274}
	
	% type= article
	\bibitem[{P.~H. {Scherrer} {et~al.}(2012){Scherrer}, {Schou}, {Bush},
		{Kosovichev}, {Bogart}, {Hoeksema}, {Liu}, {Duvall}, {Zhao}, {Title},
		{Schrijver}, {Tarbell}, \& {Tomczyk}}]{Scherrer2012}
	{Scherrer}, P.~H., {Schou}, J., {Bush}, R.~I., {et~al.} 2012,
	\bibinfo{title}{{The Helioseismic and Magnetic Imager (HMI) Investigation for
			the Solar Dynamics Observatory (SDO)},} \solphys, 275, 207,
	\dodoi{10.1007/s11207-011-9834-2}
	
	% type= book
	\bibitem[{J.~P. Serra(1982)Serra}]{Serra1982}
	Serra, J.~P. 1982, Image Analysis and Mathematical Morphology (The University
	of Michigan: Academic Press)
	
	% type= article
	\bibitem[{G.~W. {Simon} \& R.~B. {Leighton}(1964){Simon} \&
		{Leighton}}]{Simon1964}
	{Simon}, G.~W., \& {Leighton}, R.~B. 1964, \bibinfo{title}{{Velocity Fields in
			the Solar Atmosphere. III. Large-Scale Motions, the Chromospheric Network,
			and Magnetic Fields.},} \apj, 140, 1120, \dodoi{10.1086/148010}
	
	% type= article
	\bibitem[{J. {Singh} \& M.~K.~V. {Bappu}(1981){Singh} \& {Bappu}}]{Singh1981}
	{Singh}, J., \& {Bappu}, M.~K.~V. 1981, \bibinfo{title}{{A dependence on solar
			cycle of the size of the CA$^{+}$ network},} \solphys, 71, 161,
	\dodoi{10.1007/BF00153615}
	
	% type= article
	\bibitem[{J. {S{\'y}kora}(1970){S{\'y}kora}}]{Sykora1970}
	{S{\'y}kora}, J. 1970, \bibinfo{title}{{Time and Shape Changes of the
			Supergranular Network},} \solphys, 13, 292, \dodoi{10.1007/BF00153550}
	
	% type= article
	\bibitem[{S. van~der Walt {et~al.}(2014)van~der Walt, Sch{\"o}nberger,
		Nunez-Iglesias, Boulogne, Warner, Yager, Gouillart, \& Yu}]{van2014scikit}
	van~der Walt, S., Sch{\"o}nberger, J.~L., Nunez-Iglesias, J., {et~al.} 2014,
	\bibinfo{title}{scikit-image: image processing in Python,} PeerJ, 2, e453
	
	% type= article
	\bibitem[{H. {Wang}(1988){Wang}}]{Wang1988}
	{Wang}, H. 1988, \bibinfo{title}{{On the Relationship Between Magnetic Fields
			and Supergranule Velocity Fields},} \solphys, 117, 343,
	\dodoi{10.1007/BF00147252}
	
	% type= article
	\bibitem[{H. {Wang} {et~al.}(1996){Wang}, {Tang}, {Zirin}, \&
		{Wang}}]{Wang1996}
	{Wang}, H., {Tang}, F., {Zirin}, H., \& {Wang}, J. 1996, \bibinfo{title}{{The
			Velocities of Intranetwork and Network Magnetic Fields},} \solphys, 165, 223,
	\dodoi{10.1007/BF00149712}
	
	% type= article
	\bibitem[{J. {Zhao} {et~al.}(2012){Zhao}, {Couvidat}, {Bogart}, {Parchevsky},
		{Birch}, {Duvall}, {Beck}, {Kosovichev}, \& {Scherrer}}]{Zhao2012}
	{Zhao}, J., {Couvidat}, S., {Bogart}, R.~S., {et~al.} 2012,
	\bibinfo{title}{{Time-Distance Helioseismology Data-Analysis Pipeline for
			Helioseismic and Magnetic Imager Onboard Solar Dynamics Observatory (SDO/HMI)
			and Its Initial Results},} \solphys, 275, 375,
	\dodoi{10.1007/s11207-011-9757-y}
	
\end{thebibliography}

\bibliographystyle{aasjournalv7}
\newpage
\appendix
\section{Laplacian-of-Gaussian Algorithm}\label{sec:AppendixB}

The Laplacian-of-Gaussian (LoG) method applies a smoothing operator, followed by a Laplacian (second-order differential) operator, to detect blob-like structures. Before applying the LoG algorithm, the divergence map, $D$, is normalized to remove variations in the overall amplitude of the flow field as
\begin{equation}
    D_n(x,y)=\frac{D(x,y)-\langle D \rangle}{\sigma_{D}},\label{eq:log0}
\end{equation}
where $\langle D \rangle$ and $\sigma_D$ are the mean and standard deviation of the velocity divergence for each map.

The Gaussian smoothing transforms the normalized divergence map $D_n$ into a scale-space representation by convolving it with a Gaussian kernel of width $\sigma$,
\begin{equation}
\tilde f(x,y;\sigma) = G(x,y;\sigma) * D_n(x,y),\label{eq:log1}
\end{equation}
where $G(x,y;\sigma)$ is a Gaussian function with standard deviation $\sigma$, and $*$ denotes convolution. 

This multi-scale smoothing differs from Gaussian denoising (Section~\ref{sec:denoising}), which uses a fixed kernel width to suppress small-scale fluctuations. In the LoG framework, the smoothing scale $\sigma$ is a free parameter, enabling detection of features of different sizes. The Laplacian operator is then applied to the smoothed image to extract curvature information. The scale-normalized Laplacian is given by
\begin{equation}
L(x, y, \sigma) = \sigma^2\left(\frac{\partial^2 \tilde f}{\partial x^2} + \frac{\partial^2 \tilde f}{\partial y^2}\right),
\label{eq:laplacian1}
\end{equation}
which enhances blob-like structures and provides a measure of local curvature in the divergence field. Extrema of $L(x,y,\sigma)$ across spatial coordinates and scales correspond to candidate supergranular structures.

As shown by \citet{Lindeberg1998}, the characteristic radius of a detected blob is proportional to the scale parameter $\sigma$. In our implementation, $\sigma$ is expressed in pixels. The physical size of each supergranule is therefore estimated as  
\begin{equation}
r=\alpha \sigma\sqrt{2}dx, \label{eq:radius}
\end{equation}
where $\alpha=1.764$ is a correction factor that accounts for the tendency of divergence peaks to underestimate the full spatial extent of supergranular cells. The LoG detection is applied over a prescribed range of diameters representative of supergranular scales throughout their evolution, with the corresponding range of $\sigma$ values determined accordingly. The detection sensitivity is controlled by a threshold applied to the LoG response, while the allowable overlap between neighboring detections is limited to prevent excessive merging of adjacent structures. Section~\ref {sec: performance} summarizes the adopted parameters and their values.

\section{Previous studies of supergranulation properties and the magnetic activity}\label{sec:AppendixA}
*Horizontal flows are obtained from maps with local correlation tracking to obtain horizontal velocity divergence, div(U).
\begin{deluxetable}{lccccll}
\tablecaption{Reported correlations between solar activity and the size of supergranulation patterns. *Horizontal flows are obtained from maps with local correlation tracking to obtain div(U)velocity divergence.\label{tab:SG-obs}}
\tablehead{
\colhead{References} &
\colhead{Corr.} &
\colhead{Method(s)} &
\colhead{Data Type(s)} &
\colhead{Instrument} &
\colhead{Obs. dates} &
\colhead{Comments}
}
\startdata
    {\bf Non-global studies:}  &  &  &  & &  &\\
    \cite{Wang1988}      & none  & autocorrelation & Doppler & NSO/KP$^{\rm a}$& 1985 -- 1987 & few days every year\\
    \cite{Wang1996}*     & +  & autocorrelation & magnetograms & BBSO$^{\rm b}$& 1992, 1994 & few hours of observations each year, div(U) \\
    \cite{Hagenaar1997}  & none & gradient-based & CaII K &  South Pole$^{\rm c}$& 1994 &  steepest descent combined with a cluster- \\
                         &  &  &  & &  &  finding algorithm\\
    \cite{Berrilli1999}   & -- & skeletonization & CaII K & PSPT$^{\rm d}$& 1996 -- 1997 & decrease in size with latitude\\
    \cite{Raju2002}      & -- & autocorrelation & CaII K  & KSO$^{\rm e}$ & 1913 -- 1974  & used 60 images, divided by $5^{\mathrm{o}}$ in latitude\\
    \hline
    {\bf Solar Cycle variations:}     &  &  &  & &  &\\
    \cite{Sykora1970}     & +  & autocorrelation & CaII K & Arcetri Obs.$^{\rm f}$ & 1953 -- 1954,1964 & SC18; 20 observations\\
                          &    &                 &               &    & 1956 -- 1960 & \\
    \cite{Singh1981}      & -- & autocorrelation &  CaII K & KSO$^{\rm e}$ & 1907 -- 1970 & SC14 -- SC20\\
   \cite{Muenzer1989} & + & 2D FFT & CaII K & Schauinsland Obs.$^{\rm g}$& 1982 -- 1984 & SC21; 53 observations\\
   \cite{Kariyappa1994}  & -- & autocorrelation & CaII K  & KSO$^{\rm e}$ & 1958 -- 1982 & observations every 2 years; SC19 -- SC21 \\
    \cite{Komm1995}       & +  & autocorrelation & magnetograms & NSO/KP$^{\rm a}$& 1978 -- 1990 & SC21 -- SC22\\
    \cite{Meunier2003}  & + & clumping &  magnetograms & SOHO/MDI$^{\rm h}$ & 1996 -- 2002 & SC23\\
    \cite{Meunier2008}*   & -- & gradient-based & intensity & SOHO/MDI$^{\rm h}$ & 1996 -- 2006 &  SC23: 12 time series; divU \\
    \cite{McIntosh2011}   & +  & watershed & magnetograms,  & SOHO/MDI \& EIT$^{\rm h}$,  & 1996 -- 2011,  & SC23\\
                          &   &           & 304\AA  & MWO$^{\rm i}$, PSPT$^{\rm d}$, & MWO:  & SC18 -- SC20\\
                          &   &           & CaII K & STEREO/SECCHI$^{\rm j}$ & 1944 -- 1976 &\\
    \cite{Huang2012}      & -- & watershed  & magnetograms & SOHO/MDI$^{\rm g}$  & 1999 -- 2010 & SC23\\
    \cite{Chatterjee2017} & + & watershed & CaII K & KSO$^{\rm e}$, PSPT$^{\rm d}$& 1907 -- 2007 & SC14 -- SC23\\
    \cite{Mandal2017}     & +  & watershed & CaII K & KSO$^{\rm e}$ & 1907 -- 2011 & SC14 -- SC23\\
    \cite{Roudier2017}*   & none & passive tracers & Doppler     & SDO/HMI$^{\rm k}$ & 2010 -- 2015 & SC24, div(U)\\
    \cite{Rajani2022}     & -- & manual & CaII K & KSO$^{\rm e}$ & 1996 -- 2008 & SC23\\
      Current study  & -- & LoG & Doppler & SDO/HMI$^{\rm k}$ & 2010 -- 2026 & CS24 -- CS25, div(U) from\\
               &   &  &  &  &  & helioseismic inversions \\
\enddata
\end{deluxetable}
\tablenotetext{a }{NSO/KP: National Solar Observatory, Kitt Peak, USA}
\tablenotetext{b }{BBSO: Big Bear Solar Observatory, USA}
\tablenotetext{c }{Observations at South Pole by \cite{Jefferies1988}}
\tablenotetext{d }{PSPT: Precision Solar Photometric Telescope: Italy-based (1996-2016) and US-based (2005-2015)}
\tablenotetext{e }{KSO: Kodaikanal Solar Observatory, India}
\tablenotetext{f }{Osservatorio Astrofisico di Arcetri, Istituto Nazionale di AstroFisica (INAF), Italy}
\tablenotetext{g }{Kiepenheuer Institut Freiburg, Germany}
\tablenotetext{h }{SOHO: Solar and Heliospheric Observatory, MDI: Michelson Doppler Imager, EUI: Extreme-Ultraviolet Imaging Telescope}
\tablenotetext{i }{WSO: Mount Wilson Solar Observatory, USA}
\tablenotetext{j }{STEREO: Solar TErrestrial RElations Observatory, SECCHI: Sun Earth Connection Coronal and Heliospheric Investigation}
\tablenotetext{k }{SDO: Solar Dynamics Observatory, HMI: Helioseismic and Magnetic Imager}

\end{document}